\documentclass[twocolumn,twocolappendix]{aastex6}

\pdfoutput=1

\usepackage{acronym}
\usepackage{amsmath}
\usepackage{graphicx}
\usepackage{hyperref}
\usepackage{lineno}
\usepackage{multirow}

\newcommand{\svnid}[1]{ } 

\newcommand{\macro}[1]{\textcolor{red}{#1}} 

\newcommand\SECONDMONDAY{\macro{LVT151012}}

\newcommand{\Msun}{\ensuremath{\mathrm{M}_\odot}}

\newcommand{\runtime}{\macro{\ensuremath{39\,\mathrm{d}}}}
\newcommand{\OBSDAYS}{\macro{\ensuremath{\macro{16}~\mathrm{days}}}}

\newcommand{\FARone}{\macro{\ensuremath{4.9\times 10^{-6} \,\mathrm{yr}^{-1}}}}
\newcommand{\FAPone}{\macro{\ensuremath{< 2 \times 10^{-7}}}}

\newcommand{\FARtwo}{\macro{\ensuremath{0.43 \,\mathrm{yr}^{-1}}}}
\newcommand{\FAPtwo}{\macro{\ensuremath{0.02}}}

\newcommand{\MONESCOMPACT}{\macro{\ensuremath{36_{-4}^{+5}}}} 
\newcommand{\MTWOSCOMPACT}{\macro{\ensuremath{29_{-4}^{+4}}}} 
\newcommand{\REDSHIFTCOMPACT}{\macro{\ensuremath{0.09_{-0.04}^{+0.03}}}} 

\newcommand{\MONESCOMPACTSecondMonday}{\macro{\ensuremath{23_{-6}^{+18}}}} 
\newcommand{\MTWOSCOMPACTSecondMonday}{\macro{\ensuremath{13_{-5}^{+4}}}} 
\newcommand{\REDSHIFTCOMPACTSecondMonday}{\macro{\ensuremath{0.21_{-0.09}^{+0.09}}}} 

\newcommand{\ninetyrange}[3]{\ensuremath{{#1}^{+{#2}}_{-{#3}}}}
\newcommand{\ninetyinterval}[2]{\ensuremath{{#1}\text{--}{#2}}}

\newcommand{\gpcyr}{\ensuremath{\mathrm{Gpc}^3\,\mathrm{yr}}}
\newcommand{\pergpcyr}{\ensuremath{\mathrm{Gpc}^{-3}\,\mathrm{yr}^{-1}}}

\newcommand{\purekklrate}{\macro{\ensuremath{\ninetyrange{14}{39}{12}
      \, \pergpcyr}}}
\newcommand{\purekklrateinterval}{\macro{\ensuremath{\ninetyinterval{2}{53}
    \, \pergpcyr}}}

\newcommand{\threeraterangeforalltriggersundersky}{\macro{\ensuremath{\ninetyinterval{13}{600} \, \pergpcyr}}}
\newcommand{\oneraterangetorulethemallpreVTbugfix}{\macro{\ensuremath{\ninetyinterval{2}{400} \, \pergpcyr}}}
\newcommand{\oneraterangetorulethemall}{\macro{\ensuremath{\ninetyinterval{2}{600} \, \pergpcyr}}}

\newcommand{\alphabetrateonenounits}{\macro{\ensuremath{\ninetyrange{16}{38}{13}}}}
\newcommand{\alphabetratetwonounits}{\macro{\ensuremath{\ninetyrange{61}{152}{53}}}}
\newcommand{\alphabetrateoneGSTLALnoUnits}{\macro{\ensuremath{\ninetyrange{17}{39}{14}}}}
\newcommand{\alphabetratetwoGSTLALnoUnits}{\macro{\ensuremath{\ninetyrange{62}{164}{55}}}}

\newcommand{\combinedalphabetrateonenounits}{\macro{\ensuremath{\ninetyrange{17}{39}{13}}}}
\newcommand{\combinedalphabetratetwonounits}{\macro{\ensuremath{\ninetyrange{62}{165}{54}}}}

\newcommand{\alphabetratenounits}{\macro{\ensuremath{\ninetyrange{82}{155}{61}}}}
\newcommand{\alphabetrateGSTLALnoUnits}{\macro{\ensuremath{\ninetyrange{84}{172}{64}}}}
\newcommand{\combinedalphabetratenounits}{\macro{\ensuremath{\ninetyrange{83}{168}{63}}}}

\newcommand{\rateflatlognounits}{\macro{\ensuremath{\ninetyrange{63}{121}{49}}}}
\newcommand{\ratepowerlawnounits}{\macro{\ensuremath{\ninetyrange{200}{390}{160}}}}

\newcommand{\rateflatlogGSTLALnoUnits}{\macro{\ensuremath{\ninetyrange{60}{122}{48}}}}
\newcommand{\ratepowerlawGSTLALnoUnits}{\macro{\ensuremath{\ninetyrange{200}{410}{160}}}}

\newcommand{\combinedrateflatlognounits}{\macro{\ensuremath{\ninetyrange{61}{124}{48}}}}
\newcommand{\combinedratepowerlawnounits}{\macro{\ensuremath{\ninetyrange{200}{400}{160}}}}

\newcommand{\secondpforeGSTLAL}{\macro{\ensuremath{0.84}}}
\newcommand{\secondpfore}{\macro{\ensuremath{0.91}}}

\newcommand{\sensVTcentury}{\macro{\ensuremath{\ninetyrange{0.082}{0.053}{0.032}}}}

\newcommand{\ssixulone}{\ensuremath{140\,\pergpcyr}}
\newcommand{\ssixultwo}{\ensuremath{420\,\pergpcyr}}

\newcommand{\floud}{\macro{\ensuremath{0.49}}}

\renewcommand{\macro}[1]{{#1}}

\newcommand{\firstevent}{GW150914}
\newcommand{\secondevent}{\SECONDMONDAY}

\newcommand{\VTOtwolo}{\ensuremath{7}}
\newcommand{\VTOtwohi}{\ensuremath{20}}
\newcommand{\VTOthreelo}{\ensuremath{30}}
\newcommand{\VTOthreehi}{\ensuremath{70}}

\newcommand{\massone}{\macro{\ensuremath{\left(\MONESCOMPACT,\MTWOSCOMPACT\right)\,\Msun}}}
\newcommand{\masstwo}{\macro{\ensuremath{\left(\MONESCOMPACTSecondMonday,\MTWOSCOMPACTSecondMonday\right)\,\Msun}}}

\newcommand{\avgVT}{\ensuremath{\left\langle VT \right\rangle}}

\newcommand{\avgVTzero}{\ensuremath{\left\langle VT \right\rangle_0}}

\newcommand{\dd}{\ensuremath{\mathrm{d}}}
\newcommand{\diff}[2]{\ensuremath{\dfrac{\dd {#1}}{\dd {#2}}}}

\newcommand{\gstlal}{\texttt{gstlal}}
\newcommand{\pycbc}{\texttt{pycbc}}

\usepackage{color}
\usepackage{graphicx}

\newcommand{\suppseccountposterior}{4}

\begin{document}

\title{The Rate of Binary Black Hole Mergers Inferred from Advanced
  LIGO Observations Surrounding \firstevent{}}

\AuthorCallLimit=1
\fullcollaborationName{The LIGO Scientific Collaboration and Virgo
  Collaboration}
\author{%
B.~P.~Abbott,\altaffilmark{1}}  
\author{
R.~Abbott,\altaffilmark{1}  
T.~D.~Abbott,\altaffilmark{2}  
M.~R.~Abernathy,\altaffilmark{1}  
F.~Acernese,\altaffilmark{3,4}
K.~Ackley,\altaffilmark{5}  
C.~Adams,\altaffilmark{6}  
T.~Adams,\altaffilmark{7}
P.~Addesso,\altaffilmark{3}  
R.~X.~Adhikari,\altaffilmark{1}
% ...
}  

\medskip
\affiliation {$^{\dag}$Deceased, May 2015. $^{\ddag}$Deceased, March 2015.
\\
{(LIGO Scientific Collaboration and Virgo Collaboration)}%
}%
\medskip

\altaffiltext {1}{LIGO, California Institute of Technology, Pasadena, CA 91125, USA }
\altaffiltext {2}{Louisiana State University, Baton Rouge, LA 70803, USA }
\altaffiltext {3}{Universit\`a di Salerno, Fisciano, I-84084 Salerno, Italy }
\altaffiltext {4}{INFN, Sezione di Napoli, Complesso Universitario di Monte S.Angelo, I-80126 Napoli, Italy }
\altaffiltext {5}{University of Florida, Gainesville, FL 32611, USA }
\altaffiltext {6}{LIGO Livingston Observatory, Livingston, LA 70754, USA }
\altaffiltext {7}{Laboratoire d'Annecy-le-Vieux de Physique des Particules (LAPP), Universit\'e Savoie Mont Blanc, CNRS/IN2P3, F-74941 Annecy-le-Vieux, France }
\altaffiltext {8}{Albert-Einstein-Institut, Max-Planck-Institut f\"ur Gravi\-ta\-tions\-physik, D-30167 Hannover, Germany }
\altaffiltext {9}{Nikhef, Science Park, 1098 XG Amsterdam, Netherlands }
\altaffiltext {10}{LIGO, Massachusetts Institute of Technology, Cambridge, MA 02139, USA }
\altaffiltext {11}{Instituto Nacional de Pesquisas Espaciais, 12227-010 S\~{a}o Jos\'{e} dos Campos, S\~{a}o Paulo, Brazil }
\altaffiltext {12}{INFN, Gran Sasso Science Institute, I-67100 L'Aquila, Italy }
\altaffiltext {13}{INFN, Sezione di Roma Tor Vergata, I-00133 Roma, Italy }
\altaffiltext {14}{Inter-University Centre for Astronomy and Astrophysics, Pune 411007, India }
\altaffiltext {15}{International Centre for Theoretical Sciences, Tata Institute of Fundamental Research, Bangalore 560012, India }
\altaffiltext {16}{University of Wisconsin-Milwaukee, Milwaukee, WI 53201, USA }
\altaffiltext {17}{Leibniz Universit\"at Hannover, D-30167 Hannover, Germany }
\altaffiltext {18}{Universit\`a di Pisa, I-56127 Pisa, Italy }
\altaffiltext {19}{INFN, Sezione di Pisa, I-56127 Pisa, Italy }
\altaffiltext {20}{Australian National University, Canberra, Australian Capital Territory 0200, Australia }
\altaffiltext {21}{The University of Mississippi, University, MS 38677, USA }
\altaffiltext {22}{California State University Fullerton, Fullerton, CA 92831, USA }
\altaffiltext {23}{LAL, Universit\'e Paris-Sud, CNRS/IN2P3, Universit\'e Paris-Saclay, 91400 Orsay, France }
\altaffiltext {24}{Chennai Mathematical Institute, Chennai 603103, India }
\altaffiltext {25}{Universit\`a di Roma Tor Vergata, I-00133 Roma, Italy }
\altaffiltext {26}{University of Southampton, Southampton SO17 1BJ, United Kingdom }
\altaffiltext {27}{Universit\"at Hamburg, D-22761 Hamburg, Germany }
\altaffiltext {28}{INFN, Sezione di Roma, I-00185 Roma, Italy }
\altaffiltext {29}{Albert-Einstein-Institut, Max-Planck-Institut f\"ur Gravitations\-physik, D-14476 Potsdam-Golm, Germany }
\altaffiltext {30}{APC, AstroParticule et Cosmologie, Universit\'e Paris Diderot, CNRS/IN2P3, CEA/Irfu, Observatoire de Paris, Sorbonne Paris Cit\'e, F-75205 Paris Cedex 13, France }
\altaffiltext {31}{Montana State University, Bozeman, MT 59717, USA }
\altaffiltext {32}{Universit\`a di Perugia, I-06123 Perugia, Italy }
\altaffiltext {33}{INFN, Sezione di Perugia, I-06123 Perugia, Italy }
\altaffiltext {34}{European Gravitational Observatory (EGO), I-56021 Cascina, Pisa, Italy }
\altaffiltext {35}{Syracuse University, Syracuse, NY 13244, USA }
\altaffiltext {36}{SUPA, University of Glasgow, Glasgow G12 8QQ, United Kingdom }
\altaffiltext {37}{LIGO Hanford Observatory, Richland, WA 99352, USA }
\altaffiltext {38}{Wigner RCP, RMKI, H-1121 Budapest, Konkoly Thege Mikl\'os \'ut 29-33, Hungary }
\altaffiltext {39}{Columbia University, New York, NY 10027, USA }
\altaffiltext {40}{Stanford University, Stanford, CA 94305, USA }
\altaffiltext {41}{Universit\`a di Padova, Dipartimento di Fisica e Astronomia, I-35131 Padova, Italy }
\altaffiltext {42}{INFN, Sezione di Padova, I-35131 Padova, Italy }
\altaffiltext {43}{CAMK-PAN, 00-716 Warsaw, Poland }
\altaffiltext {44}{University of Birmingham, Birmingham B15 2TT, United Kingdom }
\altaffiltext {45}{Universit\`a degli Studi di Genova, I-16146 Genova, Italy }
\altaffiltext {46}{INFN, Sezione di Genova, I-16146 Genova, Italy }
\altaffiltext {47}{RRCAT, Indore MP 452013, India }
\altaffiltext {48}{Faculty of Physics, Lomonosov Moscow State University, Moscow 119991, Russia }
\altaffiltext {49}{SUPA, University of the West of Scotland, Paisley PA1 2BE, United Kingdom }
\altaffiltext {50}{University of Western Australia, Crawley, Western Australia 6009, Australia }
\altaffiltext {51}{Department of Astrophysics/IMAPP, Radboud University Nijmegen, P.O. Box 9010, 6500 GL Nijmegen, Netherlands }
\altaffiltext {52}{Artemis, Universit\'e C\^ote d'Azur, CNRS, Observatoire C\^ote d'Azur, CS 34229, Nice cedex 4, France }
\altaffiltext {53}{MTA E\"otv\"os University, ``Lendulet'' Astrophysics Research Group, Budapest 1117, Hungary }
\altaffiltext {54}{Institut de Physique de Rennes, CNRS, Universit\'e de Rennes 1, F-35042 Rennes, France }
\altaffiltext {55}{Washington State University, Pullman, WA 99164, USA }
\altaffiltext {56}{Universit\`a degli Studi di Urbino ``Carlo Bo,'' I-61029 Urbino, Italy }
\altaffiltext {57}{INFN, Sezione di Firenze, I-50019 Sesto Fiorentino, Firenze, Italy }
\altaffiltext {58}{University of Oregon, Eugene, OR 97403, USA }
\altaffiltext {59}{Laboratoire Kastler Brossel, UPMC-Sorbonne Universit\'es, CNRS, ENS-PSL Research University, Coll\`ege de France, F-75005 Paris, France }
\altaffiltext {60}{Astronomical Observatory Warsaw University, 00-478 Warsaw, Poland }
\altaffiltext {61}{VU University Amsterdam, 1081 HV Amsterdam, Netherlands }
\altaffiltext {62}{University of Maryland, College Park, MD 20742, USA }
\altaffiltext {63}{Center for Relativistic Astrophysics and School of Physics, Georgia Institute of Technology, Atlanta, GA 30332, USA }
\altaffiltext {64}{Institut Lumi\`{e}re Mati\`{e}re, Universit\'{e} de Lyon, Universit\'{e} Claude Bernard Lyon 1, UMR CNRS 5306, 69622 Villeurbanne, France }
\altaffiltext {65}{Laboratoire des Mat\'eriaux Avanc\'es (LMA), IN2P3/CNRS, Universit\'e de Lyon, F-69622 Villeurbanne, Lyon, France }
\altaffiltext {66}{Universitat de les Illes Balears, IAC3---IEEC, E-07122 Palma de Mallorca, Spain }
\altaffiltext {67}{Universit\`a di Napoli ``Federico II,'' Complesso Universitario di Monte S.Angelo, I-80126 Napoli, Italy }
\altaffiltext {68}{NASA/Goddard Space Flight Center, Greenbelt, MD 20771, USA }
\altaffiltext {69}{Canadian Institute for Theoretical Astrophysics, University of Toronto, Toronto, Ontario M5S 3H8, Canada }
\altaffiltext {70}{Tsinghua University, Beijing 100084, China }
\altaffiltext {71}{Texas Tech University, Lubbock, TX 79409, USA }
\altaffiltext {72}{The Pennsylvania State University, University Park, PA 16802, USA }
\altaffiltext {73}{National Tsing Hua University, Hsinchu City, 30013 Taiwan, Republic of China }
\altaffiltext {74}{Charles Sturt University, Wagga Wagga, New South Wales 2678, Australia }
\altaffiltext {75}{University of Chicago, Chicago, IL 60637, USA }
\altaffiltext {76}{Caltech CaRT, Pasadena, CA 91125, USA }
\altaffiltext {77}{Korea Institute of Science and Technology Information, Daejeon 305-806, Korea }
\altaffiltext {78}{Carleton College, Northfield, MN 55057, USA }
\altaffiltext {79}{Universit\`a di Roma ``La Sapienza,'' I-00185 Roma, Italy }
\altaffiltext {80}{University of Brussels, Brussels 1050, Belgium }
\altaffiltext {81}{Sonoma State University, Rohnert Park, CA 94928, USA }
\altaffiltext {82}{Northwestern University, Evanston, IL 60208, USA }
\altaffiltext {83}{University of Minnesota, Minneapolis, MN 55455, USA }
\altaffiltext {84}{The University of Melbourne, Parkville, Victoria 3010, Australia }
\altaffiltext {85}{The University of Texas Rio Grande Valley, Brownsville, TX 78520, USA }
\altaffiltext {86}{The University of Sheffield, Sheffield S10 2TN, United Kingdom }
\altaffiltext {87}{University of Sannio at Benevento, I-82100 Benevento, Italy and INFN, Sezione di Napoli, I-80100 Napoli, Italy }
\altaffiltext {88}{Montclair State University, Montclair, NJ 07043, USA }
\altaffiltext {89}{Universit\`a di Trento, Dipartimento di Fisica, I-38123 Povo, Trento, Italy }
\altaffiltext {90}{INFN, Trento Institute for Fundamental Physics and Applications, I-38123 Povo, Trento, Italy }
\altaffiltext {91}{Cardiff University, Cardiff CF24 3AA, United Kingdom }
\altaffiltext {92}{National Astronomical Observatory of Japan, 2-21-1 Osawa, Mitaka, Tokyo 181-8588, Japan }
\altaffiltext {93}{School of Mathematics, University of Edinburgh, Edinburgh EH9 3FD, United Kingdom }
\altaffiltext {94}{Indian Institute of Technology, Gandhinagar Ahmedabad Gujarat 382424, India }
\altaffiltext {95}{Institute for Plasma Research, Bhat, Gandhinagar 382428, India }
\altaffiltext {96}{University of Szeged, D\'om t\'er 9, Szeged 6720, Hungary }
\altaffiltext {97}{Embry-Riddle Aeronautical University, Prescott, AZ 86301, USA }
\altaffiltext {98}{University of Michigan, Ann Arbor, MI 48109, USA }
\altaffiltext {99}{Tata Institute of Fundamental Research, Mumbai 400005, India }
\altaffiltext {100}{American University, Washington, D.C. 20016, USA }
\altaffiltext {101}{University of Massachusetts-Amherst, Amherst, MA 01003, USA }
\altaffiltext {102}{University of Adelaide, Adelaide, South Australia 5005, Australia }
\altaffiltext {103}{West Virginia University, Morgantown, WV 26506, USA }
\altaffiltext {104}{University of Bia{\l }ystok, 15-424 Bia{\l }ystok, Poland }
\altaffiltext {105}{SUPA, University of Strathclyde, Glasgow G1 1XQ, United Kingdom }
\altaffiltext {106}{IISER-TVM, CET Campus, Trivandrum Kerala 695016, India }
\altaffiltext {107}{Institute of Applied Physics, Nizhny Novgorod, 603950, Russia }
\altaffiltext {108}{Pusan National University, Busan 609-735, Korea }
\altaffiltext {109}{Hanyang University, Seoul 133-791, Korea }
\altaffiltext {110}{NCBJ, 05-400 \'Swierk-Otwock, Poland }
\altaffiltext {111}{IM-PAN, 00-956 Warsaw, Poland }
\altaffiltext {112}{Rochester Institute of Technology, Rochester, NY 14623, USA }
\altaffiltext {113}{Monash University, Victoria 3800, Australia }
\altaffiltext {114}{Seoul National University, Seoul 151-742, Korea }
\altaffiltext {115}{University of Alabama in Huntsville, Huntsville, AL 35899, USA }
\altaffiltext {116}{ESPCI, CNRS, F-75005 Paris, France }
\altaffiltext {117}{Universit\`a di Camerino, Dipartimento di Fisica, I-62032 Camerino, Italy }
\altaffiltext {118}{Southern University and A\&M College, Baton Rouge, LA 70813, USA }
\altaffiltext {119}{College of William and Mary, Williamsburg, VA 23187, USA }
\altaffiltext {120}{Instituto de F\'\i sica Te\'orica, University Estadual Paulista/ICTP South American Institute for Fundamental Research, S\~ao Paulo SP 01140-070, Brazil }
\altaffiltext {121}{University of Cambridge, Cambridge CB2 1TN, United Kingdom }
\altaffiltext {122}{IISER-Kolkata, Mohanpur, West Bengal 741252, India }
\altaffiltext {123}{Rutherford Appleton Laboratory, HSIC, Chilton, Didcot, Oxon OX11 0QX, United Kingdom }
\altaffiltext {124}{Whitman College, 345 Boyer Avenue, Walla Walla, WA 99362 USA }
\altaffiltext {125}{National Institute for Mathematical Sciences, Daejeon 305-390, Korea }
\altaffiltext {126}{Hobart and William Smith Colleges, Geneva, NY 14456, USA }
\altaffiltext {127}{Janusz Gil Institute of Astronomy, University of Zielona G\'ora, 65-265 Zielona G\'ora, Poland }
\altaffiltext {128}{Andrews University, Berrien Springs, MI 49104, USA }
\altaffiltext {129}{Universit\`a di Siena, I-53100 Siena, Italy }
\altaffiltext {130}{Trinity University, San Antonio, TX 78212, USA }
\altaffiltext {131}{University of Washington, Seattle, WA 98195, USA }
\altaffiltext {132}{Kenyon College, Gambier, OH 43022, USA }
\altaffiltext {133}{Abilene Christian University, Abilene, TX 79699, USA }

\begin{abstract}
  A transient gravitational-wave signal, \firstevent{}, was identified
  in the twin Advanced LIGO detectors on September 14, 2015 at
  09:50:45 UTC.  To assess the implications of this discovery, the
  detectors remained in operation with unchanged configurations over a
  period of \runtime{} around the time of the signal.  At the
  detection statistic threshold corresponding to that observed for
  \firstevent{}, our search of the \OBSDAYS{} of simultaneous
  two-detector observational data is estimated to have a \ac{FAR} of
  $< \FARone{}$, yielding a $p$-value for \firstevent{} of \FAPone{}.
  Parameter estimation followup on this trigger identifies its source
  as a \ac{BBH} merger with component masses
  $(m_1, m_2) = \left(\MONESCOMPACT{}, \MTWOSCOMPACT{}\right)\, \Msun$
  at redshift $z = \REDSHIFTCOMPACT{}$ (median and 90\% credible
  range).  Here we report on the constraints these observations place
  on the rate of \ac{BBH} coalescences.  Considering only
  \firstevent{}, assuming that all \acp{BBH} in the Universe have the
  same masses and spins as this event, imposing a search \ac{FAR}
  threshold of 1 per 100 years, and assuming that the \ac{BBH} merger
  rate is constant in the comoving frame, we infer a $90\%$ credible
  range of merger rates between \purekklrateinterval{} (comoving
  frame).  Incorporating all search triggers that pass a much lower
  threshold while accounting for the uncertainty in the astrophysical
  origin of each trigger, we estimate a higher rate, ranging from
  \threeraterangeforalltriggersundersky{} depending on assumptions
  about the \ac{BBH} mass distribution.  All together, our various
  rate estimates fall in the conservative range
  \oneraterangetorulethemall{}.
\end{abstract}

\maketitle

\acrodef{BH}[BH]{black hole}
\acrodef{BBH}[BBH]{binary black hole}
\acrodef{CL}[CL]{credible level}
\acrodef{CBC}[CBC]{compact binary coalescence}
\acrodef{EOB}[EOB]{effective one body}
\acrodef{FAP}{false alarm probability}
\acrodef{FAR}{false alarm rate}
\acrodef{GRB}{gamma ray burst}
\acrodef{GW}[GW]{gravitational wave}
\acrodef{PE}{parameter estimation}
\acrodef{SNR}{signal-to-noise ratio}
\acrodefplural{SNR}{signal-to-noise ratios}

\acresetall{} 

\section{Introduction}

The first detection of a \ac{GW} signal in the twin Advanced LIGO
detectors on September 14, 2015, 09:50:45 UTC was reported in
\citet{GW150914-DETECTION}. This transient signal is designated
\firstevent{}. To assess the implications of this discovery, the
detectors remained in operation with unchanged configurations over a
period of \runtime{} around the time of the signal.  At the detection
statistic threshold corresponding to that observed for \firstevent{},
the \ac{FAR} of the search of the available \OBSDAYS{} of coincident
data is estimated to be $< \FARone{}$, yielding a $p$-value for
\firstevent{} of \FAPone{} \citep{GW150914-CBC}.  \firstevent{} is
consistent with a gravitational-wave signal from the merger of two
black holes with masses $\left(m_1, m_2\right) = \massone{}$ at
redshift $z = \REDSHIFTCOMPACT$ \citep{GW150914-PARAMESTIM}.  Here and
throughout, we report posterior medians and 90\% symmetric credible
intervals.  In this paper, we discuss inferences on the rate of
\ac{BBH} mergers from this detection and the surrounding data.  This
Letter is accompanied by \citet{RatesSupplement} (hereafter the
Supplement) containing supplementary information on our methods and
computations.

Previous estimates of the \ac{BBH} merger rate based on population
modeling are reviewed in \citet{Abadie2010}.  The range of rates given
there spans more than three orders of magnitude, from
$0.1$--$300 \, \pergpcyr$.  The rate of \ac{BBH} mergers is a crucial
output from \ac{BBH} population models, but theoretical uncertainty in
the evolution of massive stellar binaries and a lack of constraining
electromagnetic observations produce a wide range of rate estimates.
Observations of \acp{GW} can tightly constrain this rate with minimal
modeling assumptions, and thus provide useful input on the
astrophysics of massive stellar binaries.  In the absence of
detections until \firstevent{}, the most constraining rate upper
limits from \ac{GW} observations, as detailed in \citet{Aasi2013}, lie
above the model predictions.  Here, for the first time, we report on
\ac{GW} observations that constrain the model space of \ac{BBH} merger
rates.

It is possible to obtain a rough estimate of the \ac{BBH} coalescence
rate from the \firstevent{} detection by setting a low search \ac{FAR}
threshold that eliminates other search triggers \citep{GW150914-CBC}.
The inferred rate will depend on the detector sensitivity to the
\ac{BBH} population, which strongly depends on \ac{BBH} masses.
However, our single detection leaves a large uncertainty in the mass
distribution of merging \ac{BBH} systems.  \citet{Kim2003} faced a
similar situation in deriving binary neutron star merger rates from
the small sample of Galactic double neutron star systems.  They argued
that a good rate estimate follows from an approach assuming each
detected system belongs to its own class, deriving merger rates for
each class independently, and then adding the rates over classes to
infer the overall merger rate.  If we follow \citet{Kim2003}, assume
that all \ac{BBH} mergers in the universe have the same source-frame
masses and spins as \firstevent{}, and set a nominal threshold on the
search \ac{FAR} of one per century---eliminating all triggers but the
one associated with \firstevent{}---then the inferred posterior median
rate and 90\% credible range is $R_{100} = \purekklrate{}$ (see
Section \ref{sec:rates}).

Merger rates inferred from a single highly-significant trigger are
sensitive to the choice of threshold.  Less significant search
triggers eliminated under the strict \ac{FAR} threshold can also
provide information about the merger rate. For example, thresholded at
the significance of the second-most-significant trigger (designated
\secondevent{}), our search \ac{FAR} is \FARtwo{}, yielding a
$p$-value for this trigger of \FAPtwo{}.  This trigger cannot
confidently be claimed as a detection on the basis of such a
$p$-value, but neither is it obviously consistent with a terrestrial
origin, i.e.\ a result of either instrumental or environmental effects
in the detector.  Under the assumption that this trigger is
astrophysical in origin, \ac{PE} \citep{Veitch2015} indicates that its
source is also a \ac{BBH} merger with source-frame masses
$\left(m_1, m_2\right) = \masstwo{}$ at redshift
$z = \REDSHIFTCOMPACTSecondMonday{}$ \citep{GW150914-CBC}.  Based on
two different implementations of a matched-filter search, we find
posterior probabilities \secondpforeGSTLAL{} and \secondpfore{} that
\secondevent{} is of astrophysical origin (see Section
\ref{sec:counts}).  This is the only trigger besides \firstevent{}
that has probability greater than 50\% of being of astrophysical
origin.  \citet{Farr2015} presented a method by which a set of
triggers of uncertain origin like this can be used to produce a rate
estimate that is more accurate than that produced by considering only
highly significant events.

The mixture model of \citet{Farr2015} used here is similar to other
models used to estimate rates in astrophysical contexts.
\citet{Loredo1995,Loredo1998Aniso} used a similar
foreground/background mixture model to infer the rate and distribution
of \acp{GRB}.  A subsequent paper used similar models in a
cosmological context, as we do here \citep{Loredo1998Iso}.
\citet{Guglielmetti2009} used the same sort of formalism to model
ROSAT images, and it has also found use in analysis of surveys of
trans-Neptunian objects \citep{Gladman1998,Petit2008}.
\citet{Kelly2009} addresses selection effects in the presence of
population models, an issue that also appears in this work.  In
contrast to previous analyses, here we operate in the
background-dominated regime, setting a search threshold where the
\ac{FAR} is relatively high so that we can be confident that triggers
of terrestrial (as opposed to astrophysical) origin dominate near
threshold (see Section \ref{sec:counts}).

Incorporating our uncertainty about the astrophysical origin of all
search triggers that could represent \ac{BBH} signals
\citep{GW150914-CBC} using the \citet{Farr2015} method, assuming that
the \ac{BBH} merger rate is constant in comoving volume and
source-frame time, and making various assumptions about the mass
distribution of merging \ac{BBH} systems as described in Sections
\ref{sec:rates} and \ref{sec:mass-distribution}, we derive merger
rates that lie in the range $\threeraterangeforalltriggersundersky{}$.

Our rate estimates are summarized in Table \ref{tab:rate-table}; see
Section \ref{sec:rates} for more information.  Each row of Table
\ref{tab:rate-table} represents a different assumption about the
\ac{BBH} mass distribution.  The first two columns giving rates
correspond to two different search algorithms (called \pycbc{} and
\gstlal{}, described in the Supplement) with different models of the
astrophysical and terrestrial trigger distributions.  Because the rate
posteriors from the different searches are essentially identical (see
Figures \ref{fig:alphabet-posterior} and \ref{fig:rate-bounds}), the
third column giving rates provides a combined estimate that results
from an average of the posterior densities from each search.
Including the rate estimate with a strict threshold that considers
only the \firstevent{} trigger as described in Section \ref{sec:rates}
all our rate estimates lie in the conservative range
$\oneraterangetorulethemall{}$.\footnote{Following submission but
  before acceptance of this paper we identified a mistake in our
  calculation of the sensitive spacetime volume for the ``Flat'' and
  ``Power Law'' \ac{BBH} populations (see Section
  \ref{sec:mass-distribution}) that reduced those volumes and
  increased the corresponding rates by a factor of approximately two.
  Since the upper limit of this rate range is driven by the rate
  estimates for the ``Power Law'' population, the range given here
  increased when the mistake was corrected.  Previous versions of this
  paper posted to the arXiv, \citet{GW150914-DETECTION},
  \citet{GW150914-ASTRO}, and others used the mistaken rate range
  $\oneraterangetorulethemallpreVTbugfix{}$.  The correction does not
  affect the astrophysical interpretation appearing in
  \citet{GW150914-DETECTION} or \citet{GW150914-ASTRO}.}

All our rate estimates are consistent within their statistical
uncertainties, and these estimates are also consistent with the broad
range of rate predictions reviewed in \citet{Abadie2010} with only the
low end ($<1$\,\pergpcyr{}) of rate predictions being excluded.  The
astrophysical implications of the \firstevent{} detection and these
inferred rates are further discussed in \citet{GW150914-ASTRO}.

\begin{deluxetable}{lccc}
  \tablewidth{0pt}
  \tablecolumns{4} 
  \tablecaption{Rates of \ac{BBH} mergers estimated under various
    assumptions.  See Section \ref{sec:rates}.  All results are
    reported as a posterior median and 90\% symmetric credible
    interval.
    \label{tab:rate-table}}
  \tablehead{ Mass Distribution & \multicolumn{3}{c}{$R / \left(\pergpcyr{}\right)$} \\ & \pycbc{} & \gstlal{} & Combined }
  \startdata
  \firstevent{} & \alphabetrateonenounits{} & \alphabetrateoneGSTLALnoUnits{} & \combinedalphabetrateonenounits{} \\
  \secondevent{} & \alphabetratetwonounits{} & \alphabetratetwoGSTLALnoUnits{} & \combinedalphabetratetwonounits{} \\
  Both & \alphabetratenounits{} & \alphabetrateGSTLALnoUnits{} & \combinedalphabetratenounits{} \\
  \cutinhead{Astrophysical}
  Flat in log mass & \rateflatlognounits{} & \rateflatlogGSTLALnoUnits{} & \combinedrateflatlognounits{} \\
  Power Law (-2.35) & \ratepowerlawnounits{} & \ratepowerlawGSTLALnoUnits{} & \combinedratepowerlawnounits{} \\
  \enddata
\end{deluxetable}

This letter presents the results of our rate inference.  For
methodological and other details of the analysis, see the Supplement.

The results presented here depend on assumptions about the masses,
spins and cosmological distribution of sources. As \ac{GW} detectors
acquire additional data and their sensitivities improve, we will be
able to test these assumptions and deepen our understanding of
\ac{BBH} formation and evolution in the Universe.

\section{Rate Inference}
\label{sec:rate-inference}

A rate estimate requires counting the number of signals in an
experiment and then estimating the sensitivity to a population of
sources to transform the count into an astrophysical rate.
Individually, the count of signals and the sensitivity will depend on
specific detection and trigger generation thresholds imposed by the
pipeline, but the estimated rates should not depend strongly on such
thresholds.  We consider various methods of counting signals, employ
two distinct search pipelines and obtain a range of broadly consistent
rate estimates.

\subsection{Counting Signals}
\label{sec:counts}

Two independent pipelines searched the coincident data for signals
matching a \ac{CBC} \citep{GW150914-CBC}, each producing a set of
coincident search triggers.  Both the \pycbc{} pipeline
\citep{Usman2015} and the \gstlal{} pipeline
\citep{gstlalmethods:2015} perform matched-filter searches for
\ac{CBC} signals using aligned-spin templates
\citep{Taracchini:2013rva,Purrer2015} when searching the \ac{BBH}
parts of the \ac{CBC} parameter space.  In these searches,
single-detector triggers are recorded at maxima of the \ac{SNR} time
series for each template \citep{Allen:2005fk}; coincident search
triggers are formed when pairs of triggers, one from each detector,
occur in the same template with a time difference of
$\pm15\,\mathrm{ms}$ or less.  Our data set here consists of the set
of coincident triggers returned by each search over the \OBSDAYS{} of
coincident observations.  See the Supplement for more information
about the generation of triggers.

The \citet{Farr2015} framework considers two classes of coincident
triggers: those whose origin is astrophysical and those whose origin
is terrestrial.  Terrestrial triggers are the result of either
instrumental or environmental effects in the detector.  The two types
of sources produce triggers with different densities in the space of
detection statistics, which we denote as $x$.  We consider all
triggers above a threshold chosen so that triggers of terrestrial
origin dominate at the threshold.  Triggers appear above threshold in
a Poisson process with number density in detection space
\begin{equation}
  \label{eq:two-pop-fgmc-rate}
  \diff{N}{x} = \Lambda_1 p_1(x) + \Lambda_0 p_0(x),
\end{equation}
where the subscripts ``1'' and ``0'' refer to the astrophysical and
terrestrial origin, $\Lambda_1$ and $\Lambda_0$ are the Poisson mean
numbers of triggers of astrophysical and terrestrial type, and $p_1$
and $p_0$ are the (normalized) density of triggers of astrophysical
and terrestrial origin over detection space.  We estimate the
densities, $p_0$ and $p_1$, of triggers of the two types empirically
as described in the Supplement and in \citet{GW150914-CBC}.  Here we
ignore the time of arrival of the triggers in our data set, averaging
the rates of each type of trigger and the sensitivity of the detector
to astrophysical signals over time.  We do this because it is
difficult to estimate $p_0$ and $p_1$ over short times and because we
see no evidence of time variation in $p_0$ and $p_1$; for more details
see the Supplement.

The parameter $\Lambda_1$ is the mean number of signals of
astrophysical origin above the chosen threshold; it is not the mean
number of signals confidently detected (see Section
\ref{sec:discussion}).  Under the assumptions we make here of a rate
that is constant in the comoving frame, $\Lambda_1$ is related to the
astrophysical rate of \ac{BBH} coalescences $R$ by
\begin{equation}
  \label{eq:rate-count-relation}
  \Lambda_1 = R \avgVT,
\end{equation}
where $\left\langle VT \right\rangle$ is the time- and
population-averaged spacetime volume to which the detector is
sensitive at the chosen search threshold, defined in Eq.\
\eqref{eq:average-space-time-volume}.  Because the astrophysical rate
enters the likelihood only in the combination $R \avgVT$, which
represents a dimensionless count, we first discuss estimation of
$\Lambda$ in this Section, and then discuss the relationship between
the posterior on $\Lambda$ and on the rate $R$ in Section
\ref{sec:rates}.  

The likelihood for a trigger set with detection statistics
$\left\{ x_j | j = 1, \ldots, M \right\}$ is
\citep{Loredo1995,Farr2015}
\begin{multline}
  \label{eq:two-pop-likelihood}
  \mathcal{L}\left(\left\{ x_j | j = 1, \ldots, M\right\} | \Lambda_1,
    \Lambda_0 \right) \\ = \left\{\prod_{j = 1}^{M} \left[ \Lambda_1 p_1\left(x_j\right) + \Lambda_0 p_0\left( x_j \right) \right] \right\} \exp\left[ -\Lambda_1 - \Lambda_0 \right].
\end{multline}
See the Supplement for a derivation of this likelihood function for
our Poisson mixture model.  In each pipeline, the \emph{shape} of the
astrophysical trigger distribution $p_1(x)$ is, to a very great
extent, \emph{universal} \citep{Schutz2011,Chen2014}; that is, it does
not depend on the properties of the source (see Supplement).  It is
this remarkable property that motivates this approach to our analysis.
In principle the data from the LIGO detector contain much more
information than can be summarised by a trigger with detection
statistic $x$.  For example, to obtain information about the source
associated to a trigger (if any) we can follow it up with a separate
\ac{PE} analysis \citep{Veitch2015}.  Unfortunately, with only two
likely astrophysical sources\footnote{Actually three, as this article
  goes to press \citep{Abbott2016GW151226,Abbott2016O1BBH}.}, the
amount of information available about the distribution of source
properties is minimal.  Since we cannot eliminate the dominant
astrophysical systematic of uncertainty about the distribution of
source parameters through a more detailed analysis, we here adopt this
simpler method.  In the future, as detections accumulate, we expect to
transition to a method of analysis that incorporates estimation of
population parameters.  Here we deal with the uncertainty in the
astrophysical population by estimating rates under several different
assumptions about the population; see Section \ref{sec:rates}.

We impose a prior on the $\Lambda$ parameters of:
\begin{equation}
  \label{eq:two-pop-prior}
  p\left(\Lambda_1,\Lambda_0\right) \propto
    \frac{1}{\sqrt{\Lambda_1}} \frac{1}{\sqrt{\Lambda_0}}.
\end{equation}
See Section \suppseccountposterior{} of the Supplement for a
discussion of our choice of prior.  The posterior on expected counts
is proportional to the product of the likelihood from Eq.\
\eqref{eq:two-pop-likelihood} and the prior from Eq.\
\eqref{eq:two-pop-prior}:
\begin{multline}
  \label{eq:two-pop-posterior}
  p\left(\Lambda_1, \Lambda_0 | \left\{ x_j | j = 1, \ldots, M\right\}
  \right) \\ \propto \left\{ \prod_{j = 1}^{M} \left[ \Lambda_1
    p_1\left(x_j\right) + \Lambda_0 p_0\left( x_j \right) \right] \right\}
  \\ \times \exp\left[ -\Lambda_1 - \Lambda_0 \right]
  \frac{1}{\sqrt{\Lambda_1 \Lambda_0}}.
\end{multline}
Posterior distributions for $\Lambda_0$ and $\Lambda_1$ were obtained
using a Markov-chain Monte Carlo; details are given in the Supplement,
along with the resulting expected counts $\Lambda_1$.

Using the posterior on the $\Lambda_0$ and $\Lambda_1$, we can compute
the posterior probability that each particular trigger comes from an
astrophysical versus terrestrial source.  The conditional probability
that an event at detection statistic $x$ comes from an astrophysical
source is given by \citep{Guglielmetti2009,Farr2015}
\begin{equation}
  \label{eq:pfore-conditional}
  P_1\left(x|\Lambda_0, \Lambda_1\right) = \frac{\Lambda_1 p_1(x)}{\Lambda_1 p_1(x) + \Lambda_0 p_0(x)}.
\end{equation}
Marginalizing over the posterior for the expected counts gives
\begin{multline}
  \label{eq:pfore}
  P_1\left(x | \left\{ x_j | j = 1, \ldots, M\right\}\right) \\ \equiv
  \int \dd \Lambda_0 \, \dd \Lambda_1 \, P_1\left(x|\Lambda_0,
    \Lambda_1\right) \\ \times p\left(\Lambda_1, \Lambda_0 | \left\{
      x_j | j = 1, \ldots, M\right\} \right),
\end{multline}
which is the posterior probability that an event at detection
statistic $x$ is astrophysical in origin given the observed event set
(and associated count inference).  In particular, we calculate the
posterior probability that \secondevent{} is of astrophysical origin
to be \secondpforeGSTLAL{} with the \gstlal{} pipeline and
\secondpfore{} with the \pycbc{} pipeline.  These probabilities, while
not high enough to claim \secondevent{} as a second detection, are
large enough to motivate exploring a second class of \ac{BBH}s in the
\citet{Kim2003} prescription.

It is more difficult to estimate the posterior probability that
\firstevent{} is of astrophysical origin because there are no samples
from the empirical background estimation in this region, so the
probability estimate is sensitive to how the background density,
$p_0$, is analytically extended into this region.  We estimate that
the probability of astrophysical origin for \firstevent{} is larger
than $1 - 10^{-6}$.

Under the assumption that \firstevent{} and \secondevent{} are
astrophysical, posterior distributions for system parameters can be
derived \citep{Veitch2015}.  Both triggers are consistent with
\ac{BBH} merger sources with masses $\left(m_1,m_2\right) = \massone$
at redshift $\REDSHIFTCOMPACT$ (\firstevent{}) and
$\left(m_1,m_2\right) = \masstwo$ at redshift
$\REDSHIFTCOMPACTSecondMonday$ (\secondevent{})
\citep{GW150914-PARAMESTIM,GW150914-CBC}.  Following \citet{Kim2003},
we consider the second event, if astrophysical, to be a separate class
of \ac{BBH} from \firstevent{}.

We can incorporate a second class of \ac{BBH} merger into our mixture
model in a straightforward way.  Let there be \emph{two} classes of
\ac{BBH} mergers: type 1, which are \firstevent{}-like and type 2,
which are \secondevent{}-like.  Our trigger set then consists of
triggers of type 1, triggers of type 2, and triggers of terrestrial
origin, denoted as type 0.  The distribution of triggers over
detection statistic, $x$, now follows an inhomogeneous Poisson process
with three terms:
\begin{equation}
  \label{eq:rate}
  \diff{N}{x} = \Lambda_1 p_1(x) + \Lambda_2 p_2(x) + \Lambda_0 p_0(x),
\end{equation}
where $\Lambda_1$, $\Lambda_2$, and $\Lambda_0$ are the mean number of
triggers of each type in the data set, and $p_1$, $p_2$, and $p_0$ are
the probability densities for triggers of each type over the detection
statistic.  The shape of the distribution of \acp{SNR} is independent
of the event properties, so $p_1 = p_2$ \citep{Schutz2011,Chen2014};
we cannot distinguish \ac{BBH} classes based only on their detection
statistic distributions, but rather require \ac{PE}.

When an event's parameters are known to come from a certain class $i$,
under the astrophysical origin assumption, then the trigger rate
becomes
\begin{equation}
  \label{eq:rate-i}
  \diff{N_i}{x} = \Lambda_i p_i(x) + \Lambda_0 p_0(x),
\end{equation}
i.e.\ we permit the event to belong to either its astrophysical class
or to an terrestrial source, but not to the other astrophysical
class.
The Poisson likelihood for the set of $M$ triggers,
$\left\{ x_j | j = 1, \ldots, M \right\}$, exceeding our detection
statistic threshold is similar to Eq.~\eqref{eq:two-pop-likelihood},
but we now account for the distinct classification of \firstevent{}
and \secondevent{} based on \ac{PE}:
\begin{multline}
  \label{eq:likelihood}
  \mathcal{L}\left(\left\{ x_j | j = 1, \ldots, M\right\} | \Lambda_1,
    \Lambda_2, \Lambda_0 \right) \\ = \left[ \Lambda_1
    p_1\left(x_1\right) + \Lambda_0 p_0\left(x_1\right)\right]
  \left[\Lambda_2 p_2 \left( x_2 \right) + \Lambda_0 p_0
    \left(x_2\right) \right] \\ \times \left\{ \prod_{j = 3}^{M} \left[
    \Lambda_1 p_1\left(x_j\right) + \Lambda_2 p_2\left( x_j \right) +
    \Lambda_0 p_0\left( x_j \right) \right] \right\} \\ \times \exp\left[
    -\Lambda_1 - \Lambda_2 - \Lambda_0 \right].
\end{multline}
The first two terms in this product are the rates of the form of 
Eq.~\eqref{eq:rate-i} for the \firstevent{} and \secondevent{} triggers,
whose class, if not terrestrial, is known; the remaining terms in the
product over coincident triggers represent the other events, whose
class is not known.  

As above, the expected counts of type 1 and 2 triggers are related to
the astrophysical rates of the corresponding events by
\begin{equation}
  \label{eq:two-pop-rate}
  \Lambda_i = R_i \avgVT_i,
\end{equation}
where $\avgVT_i$ is the time- and population-averaged space-time
volume to which the detector is sensitive for event class $i$, defined
in Eq.\ \eqref{eq:average-space-time-volume} under the population
assumption in Eq.\ \eqref{eq:delta-population}.

We impose a prior for the total astrophysical and terrestrial expected
counts:
\begin{equation}
  \label{eq:prior}
  p\left(\Lambda_1,\Lambda_2,\Lambda_0\right) \propto
    \frac{1}{\sqrt{\Lambda_1 + \Lambda_2}} \frac{1}{\sqrt{\Lambda_0}}
\end{equation}
This prior is chosen to match the prior in Eq.\
\eqref{eq:two-pop-prior}.  It is adequate for the two events that we
are analysing here, but should be modified if a large number of events
are being analysed in this formalism; with $N$ categories of
foreground event under this prior, the expected number of total counts
becomes $N + 1/2$.  The posterior on expected counts given the trigger
set is proportional to the product of likelihood, Eq.\
\eqref{eq:likelihood}, and prior, Eq.\ \eqref{eq:prior}:
\begin{multline}
  \label{eq:count-posterior}
  p\left( \Lambda_1, \Lambda_2, \Lambda_0 | \left\{ x_j \right\}
  \right) \\ \propto \mathcal{L}\left(\left\{ x_j \right\} | \Lambda_1,
    \Lambda_2, \Lambda_0 \right) p\left(\Lambda_1,\Lambda_2,\Lambda_0\right)
\end{multline}
We again use Markov-chain Monte Carlo samplers to obtain resulting
expected counts for $\Lambda_1$, $\Lambda_2$, and
$\Lambda \equiv \Lambda_1 + \Lambda_2$.  These parameters represent
the Poisson mean number of events of type 1 (\firstevent{}-like), type
2 (\secondevent{}-like), and both types over the observation, above a
very low detection statistic threshold.  The estimates, which are
given in the Supplement, are consistent with one event of
astrophysical origin (\firstevent{}) at very high probability, a
further trigger (\secondevent{}) with high probability, and possibly
several more of each type in the set of triggers at lower
significance.  In the next subsection, we will describe how to turn
these expected counts of events into astrophysical rates.

\subsection{Rates}
\label{sec:rates}

The crucial element in the step from expected counts to rates is to
determine the sensitivity of the search.  Search sensitivity is
described by the selection function, which gives, as a function of
source parameters, the probability of generating a trigger above the
chosen threshold.  Here we assume that events are uniformly
distributed in comoving volume and source time, and describe the
distribution of the other parameters (masses, spins, orientation
angles, etc., here denoted by $\theta$) for events of type $i$ by a
distribution function $s_i(\theta)$.  Because the shape of the
distribution $p_1(x)$ is universal \citep{Schutz2011,Chen2014}, the
source population enters the likelihood only through the search
sensitivity; this situation differs from previous astrophysical rate
calculations
\citep{Loredo1995,Loredo1998Iso,Loredo1998Aniso,Gladman1998}, where
information about the source properties is contained in each trigger.
Under these assumptions, a count at a chosen threshold $\Lambda_i$ is
related to an astrophysical rate $R_i$ by
\begin{equation}
  \label{eq:count-rate}
  \Lambda_i = R_i \avgVT_i,
\end{equation}
where
\begin{equation}
  \label{eq:average-space-time-volume}
  \avgVT_i = T \int \dd z \, \dd \theta \, \diff{V_c}{z} \frac{1}{1+z} s_i(\theta) f(z,\theta)
\end{equation}
(see Supplement).  Here $R_i$ is the space-time rate density in the
comoving frame, $0 \leq f(z, \theta) \leq 1$ is the selection
function, $T$ is the total observation time in the observer frame, and
$V_c(z)$ is the comoving volume contained within a sphere out to
redshift $z$ \citep{Hogg1999}.%
\footnote{Throughout this paper, we use the ``TT+lowP+lensing+ext''
  cosmological parameters from Table 4 of \citet{Planck2015}.} %
In other words, the posterior on $R_i$ is obtained by substituting
Eq.\ \eqref{eq:count-rate} into Eq.\ \eqref{eq:count-posterior}.
We need to know (or assume) $s_i$, the population distribution for
events of type $i$, before we can turn expected counts into rates.

The \citet{Kim2003} assumption is that the population follows the
observed sources:
\begin{equation}
  \label{eq:delta-population}
  s_i(\theta) = \delta\left( \theta - \theta_i \right),
\end{equation}
where $\delta$ is the Dirac delta function and $\theta_i$ are the
parameters of source type $i$.  Because of the finite SNR of the
events, we do not know these parameters perfectly; we marginalize over
our imperfect knowledge by integrating over the \ac{PE} posterior for
the intrinsic, source-frame parameters from the followup on each
trigger.  This effectively replaces the Dirac delta in Eq.\
\eqref{eq:delta-population} by the \ac{PE} posterior distribution in
the integral in Eq.\ \eqref{eq:average-space-time-volume}.

\subsubsection{Rate Using \firstevent{} Only}

It is possible to obtain a rough estimate of the \ac{BBH} coalescence
rate from only the \firstevent{} trigger using a high-significance
threshold on the \ac{FAP}, and a correspondingly restrictive selection
function to estimate sensitive time-volumes, so that only this one
event is above threshold.  We impose a nominal one-per-century
threshold on the search \ac{FAR}.  The \firstevent{} trigger is the
only one that exceeds this threshold.  We estimate the integral in
Eq.\ \eqref{eq:count-rate} via a Monte Carlo procedure.  We add
simulated \ac{BBH} signals to the detector data streams with
source-frame masses and spins sampled from the posterior distributions
from the \ac{PE} of the \firstevent{} trigger described in
\citet{GW150914-PARAMESTIM,GW150914-CBC}, random orientations and sky
locations, and a fixed redshift distribution.%
\footnote{The source- and observer-frame masses are related by a
  redshift factor, $(1+z) M^\mathrm{source} = M^\mathrm{observer}$.} %
The waveforms correspond to \ac{BBH} systems with spins aligned with
the orbital angular momentum and are generated using the \ac{EOB}
formalism \citep{Taracchini:2013rva,Purrer2015}; in nature we would
never expect perfect spin alignment, but nevertheless the \ac{EOB}
waveforms are a good fit to the observed data
\citep{GW150914-PARAMESTIM} and we therefore expect they will
accurately represent our true detection efficiency for sources of this
type.

We search this modified data stream and record all injections found
above the threshold for inclusion in our trigger sets.  By weighting
the recovered injections appropriately, we can estimate the integral
in Eq.\ \eqref{eq:average-space-time-volume}, and, accounting for the
effect on the recovered luminosity distance from a 10\% amplitude
calibration uncertainty \citep{GW150914-CALIBRATION}, we obtain
$\avgVT_{100} = \sensVTcentury{} \, \gpcyr$.  See the Supplement
for details.  Systematic
uncertainties in the waveforms used for the injections and search are
estimated to induce an uncertainty in the sensitive volume calculation
that is much smaller than the calibration uncertainty
\citep{Aasi2013,Littenberg2013}.  

With such a high threshold, any trigger is virtually certain to be
astrophysical in origin, so $p_0 \simeq 0$ (see Section
\ref{sec:counts}), thus the posterior on the associated rate,
$R_{100}$ becomes:
\begin{multline}
  \label{eq:pure-kkl-posterior}
  p\left(R_{100}|\mathrm{\firstevent{}}\right) \propto \\
  \sqrt{R_{100} \avgVT_{100}}
  \exp\left[ - R_{100} \avgVT_{100} \right],
\end{multline}
from which we infer $R_{100} = \purekklrate{}$.

\subsubsection{Rates Incorporating All Triggers}

As discussed in Section~\ref{sec:counts}, there is useful information
about the merger rate from triggers with \ac{FAR} less significant
than one per century.  Following \citet{Farr2015} we set a lower
acceptance threshold such that the trigger density at threshold is
dominated by triggers of terrestrial origin.  As before, we perform a
Monte Carlo estimation of the integral in
Eq.~\eqref{eq:average-space-time-volume} using posterior distributions
from the \ac{PE} of both the \firstevent{} and \secondevent{}
described in \citet{GW150914-PARAMESTIM,GW150914-CBC}, but with the lower thresholds
used in Section~\ref{sec:counts}; the results are given in
the Supplement.  

Figure \ref{fig:alphabet-posterior} shows the posterior we infer on
the rates $R_1$, $R_2$, and $R \equiv R_1 + R_2$ from our estimates of
$\avgVT{}_{1,2}$ and the posteriors on the expected counts from
Section \ref{sec:counts}.  Results are shown in
Table~\ref{tab:rate-table} in the rows \firstevent{}, \secondevent{},
and Both.  Because the two pipelines give rate estimates that are in
excellent agreement with each other, we report a combined rate that
gives the median and 90\% symmetric credible range for a posterior
that is the average of the posterior derived from each pipeline
independently.  Here, $R_1$ and $R_2$ are the contributions to the
rate from systems of each class, and $R$ should be interpreted as the
total rate of \ac{BBH} mergers in the local universe.

\begin{figure}
  \plotone{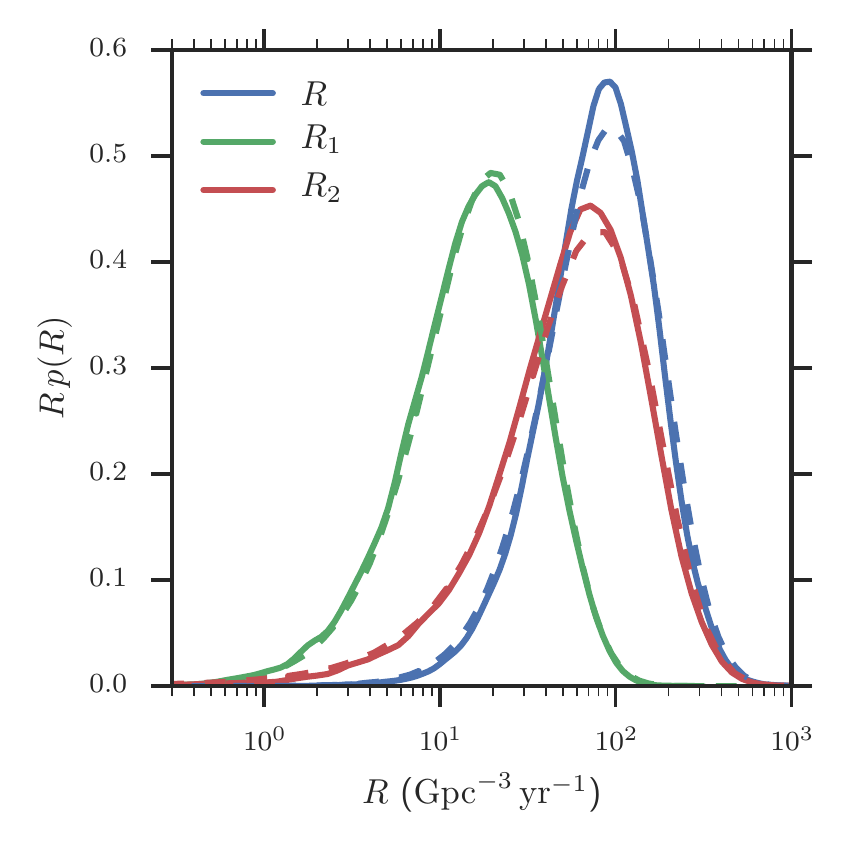}
  \caption{The posterior density on the rate of \firstevent{}-like
    \ac{BBH} inspirals, $R_1$ (green), \secondevent{}-like \ac{BBH}
    inspirals, $R_2$ (red), and the inferred total rate,
    $R = R_1 + R_2$ (blue).  The median and 90\% credible levels are
    given in Table~\ref{tab:rate-table}.  Solid lines give the rate
    inferred from the \pycbc{} trigger set, while dashed lines give
    the rate inferred from the \gstlal{} trigger set.}
  \label{fig:alphabet-posterior}
\end{figure}

\section{Sensitivity to Astrophysical Mass Distribution}
\label{sec:mass-distribution}

The assumptions in the \citet{Kim2003} method about the distribution
of intrinsic \ac{BBH} population parameters are strong and almost
certainly unrealistic.  To test the sensitivity of our rate estimate
to assumptions about \ac{BH} masses, we report in this section on two
additional estimates of the rate using different source distributions
$s(\theta)$ that bracket possible astrophysical scenarios.  

The first source distribution we take to have spins aligned with the
orbital angular momentum, with magnitude uniform in
$-0.99 \leq (a/m)_{1,2} \leq 0.99$ and masses flat in
$\log\left(m_1\right)$ and $\log\left(m_2\right)$,
\begin{equation}
  \label{eq:flat-in-log-distribution}
  s(\theta) \sim \frac{1}{m_1} \frac{1}{m_2},
\end{equation}
with $m_1, m_2 \geq 5\,\Msun$ and $m_1 + m_2 \leq 100\,\Msun$.  The
spin distribution of merging \acp{BH} is very uncertain, but is
unlikely to be concentrated at $a = 0$, so a uniform distribution of
$a$ is a reasonable choice.  The leading-order term in the \ac{GW}
amplitude depends only on the masses of the system, so our results are
not particularly sensitive to the choice of spin distribution.  This
flat distribution in mass probably weights more heavily toward
high-mass \acp{BH} than the true astrophysical distribution
\citep{Fryer2001,Fryer2012,Dominik2012,Spera2015}.  Coalescences of
higher-mass \acp{BH} from the $5\,\Msun$--$100\,\Msun$ range produce
higher signal-to-noise ratios in the detectors at the same distance,
so this time-volume estimate is probably higher than that for the true
astrophysical distribution; the corresponding rate estimate is
therefore probably lower than the true \ac{BBH} rate.  We choose
$5\,\Msun$ for the lower mass limit because it encompasses the
inferred mass range from \ac{PE} on \secondevent{} and because there
are indications of a mass gap between the heaviest neutron stars and
the lightest \acp{BH} \citep{Ozel2010,Farr2011}; but see
\citet{Kreidberg2012} for an alternative explanation for the dearth of
\ac{BH} mass estimates below $\sim 5 \, \Msun$.  Using an injection
campaign as described above, and incorporating calibration
uncertainty, we estimate the sensitive time-volume for this
population; the results are given in the Supplement.

The second source distribution we take to have the same spin
distribution with masses following a power-law on the larger BH
mass,\footnote{The power chosen here is the same as the Salpeter
  initial mass function \citep{Salpeter1955}, but this should not be
  understood to suggest that the distribution of the more massive
  \ac{BH} in a binary would follow the IMF; the initial mass--final
  mass relation for massive stars is complicated and nonlinear
  \citep{Fryer2001,Fryer2012,Dominik2012,Spera2015}.  Instead, as
  described in the text, this distribution is designed to provide a
  reasonable lower-limit for the sensitive time-volume and upper limit
  for the rate.}
\begin{equation}
  \label{eq:power-law-larger-mass}
  p\left( m_1 \right) \sim m_1^{-2.35},
\end{equation}
with the smaller mass distributed uniformly in $q \equiv m_2/m_1$, and
with $m_1, m_2 \geq 5 \, \Msun$ and $m_1 + m_2 \leq 100 \, \Msun$.
The results of using this distribution in an injection campaign are
given in the Supplement.
This distribution likely produces more low-mass \acp{BH} than the true
astrophysical distribution, and therefore the sensitive time-volume is
probably smaller than would be obtained with the true distribution;
the estimated rate is correspondingly higher
\citep{Fryer2001,Fryer2012,Dominik2012,Spera2015}.

\begin{figure}
  \plotone{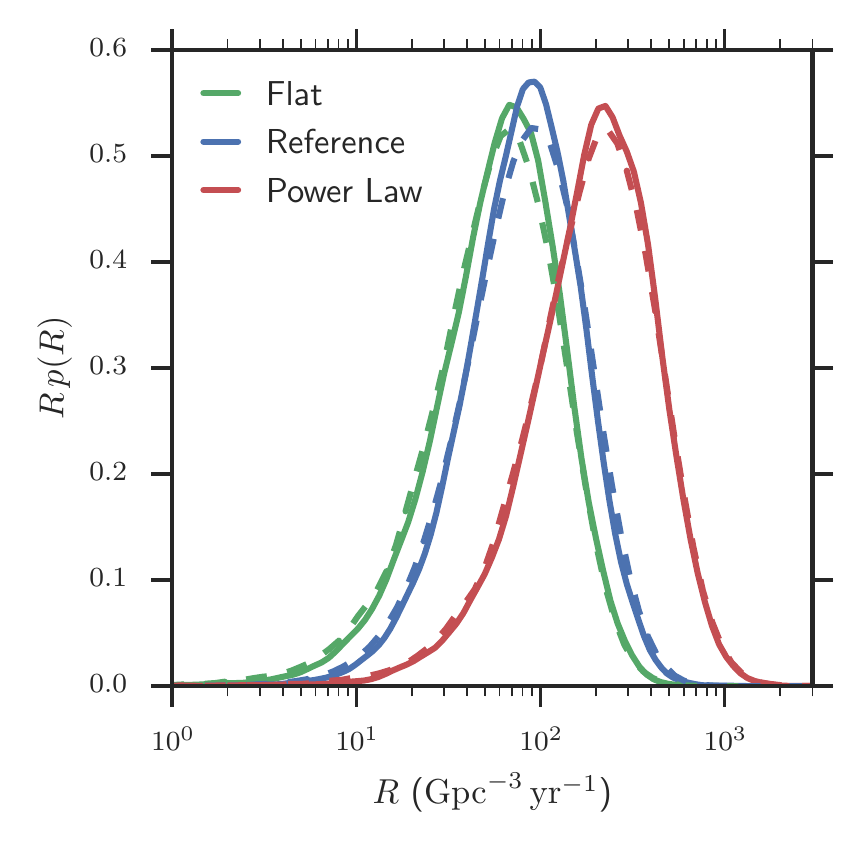}
  \caption{Sensitivity of the inferred \ac{BBH} coalescence rate to
    the assumed astrophysical distribution of \ac{BBH} masses.  The
    curves represent the posterior assuming that \ac{BBH} masses are
    flat in $\log\left(m_1\right)$--$\log\left(m_2\right)$, as in Eq.\
    \eqref{eq:flat-in-log-distribution} (green; ``Flat''), are exactly
    \firstevent{}-like or \secondevent{}-like as described in Section
    \ref{sec:rate-inference} (blue; ``Reference''), or are
    distributed as in Eq.\ \eqref{eq:power-law-larger-mass} (red;
    ``Power Law'').  The \pycbc{} results are shown in solid lines and
    the \gstlal{} results are shown in dotted lines.  Though the
    searches differ in their models of the astrophysical and
    terrestrial triggers, the rates inferred from each search are very
    similar.  The posterior median rates and symmetric 90\% \ac{CL}
    intervals are given in Table~\ref{tab:rate-table}.  Comparing to
    the total rate computed using the assumptions in \citet{Kim2003}
    in Section \ref{sec:rate-inference} we see that the rate can
    change by a factor of a few depending on the assumed \ac{BBH}
    population.  In spite of this seemingly-large variation, all three
    rate posterior distributions are consistent within our statistical
    uncertainties.
  \label{fig:rate-bounds}
}
\end{figure}

We use the same astrophysical and terrestrial trigger densities as
described in Section \ref{sec:counts}; we now consider all triggers to
belong to only two populations, an astrophysical and a terrestrial
population, as in the analysis at the beginning of Section
\ref{sec:counts} (see Eq.\ \eqref{eq:two-pop-posterior}).  We relate
expected counts $\Lambda_1$ to rates via Eq.~\eqref{eq:count-rate},
with the $\avgVT$ for the astrophysical distributions given in the
Supplement.  We find
$R_\mathrm{flat} = \combinedrateflatlognounits{} \, \pergpcyr$ and
$R_\mathrm{pl} = \combinedratepowerlawnounits{} \, \pergpcyr$.
Posteriors on the rates, together with the reference \ac{BBH}
coalescence rate $R$ from Section \ref{sec:rate-inference}, appear in
Figure \ref{fig:rate-bounds}.  A summary of the various inferred rates
appears in Table \ref{tab:rate-table}.

\section{Discussion}
\label{sec:discussion}

In the absence of clear detections, previous LIGO-Virgo observing runs
have yielded merger rate upper limits \citep{Aasi2013}.  Even the most
optimistic assumptions about the \ac{BBH} distribution from Section
\ref{sec:mass-distribution} imply rates that are significantly below
the rate upper limits for the same distribution of masses implied by
the results of \citet{Aasi2013}.  For the rate estimates from Section
\ref{sec:rates}, the corresponding upper limits from \citet{Aasi2013}
are $\ssixulone{}$ for \firstevent{}-like systems and $\ssixultwo{}$
for \secondevent{}-like systems; compared to
$R_1 = \combinedalphabetrateonenounits{}\,\pergpcyr$ and
$R_2 = \combinedalphabetratetwonounits{}\,\pergpcyr$, it is clear that
the sensitive time-volume reach of Advanced LIGO, even from only
\OBSDAYS{} of coincident observations, is vastly larger than that of
any previous gravitational-wave observations.

The search thresholds used in our analysis are much smaller than those
required to produce a confident detection.  We estimate that a
fraction \floud{} of the events exceeding our search threshold in the
\pycbc{} pipeline would also exceed a one-per-century \ac{FAR}
threshold.  One may wonder, then, how many of these significant events
we can expect to see in future observations.

\begin{figure*}
  \plottwo{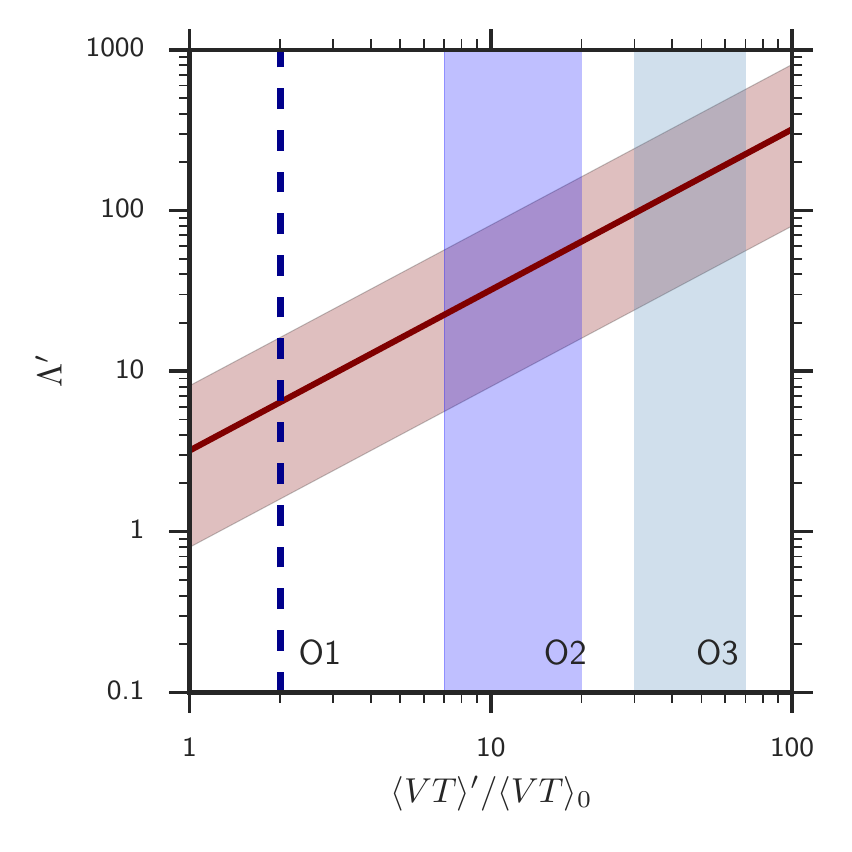}{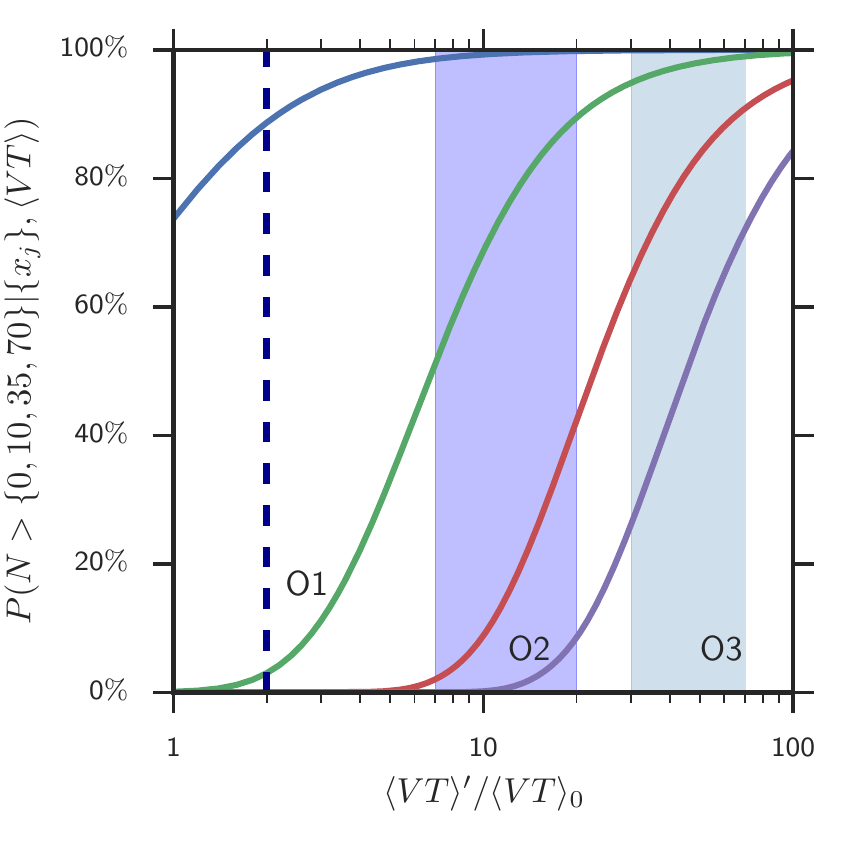}
  \caption{Left panel: The median value and $90\%$ credible interval
    for the expected number of highly significant events (\acp{FAR}
    $<$1/century) as a function of surveyed time-volume in an
    observation (shown as a multiple of $\avgVTzero$). The expected
    range of values of $\avgVT{}$ for the observations in O2 and O3
    are shown as vertical bands. Right panel: The probability of
    observing $N>0$ (blue), $N>10$ (green), $N>35$ (red), and $N>70$
    (purple) highly significant events, as a function of surveyed
    time-volume. The vertical line and bands show, from left to right,
    the expected sensitive time-volume for each of the O1 (dashed
    line), O2, and O3 observations.}
 \label{fig:lprime}
 \end{figure*}

For a Poisson mean occurrence number $\Lambda$ in an experiment with
sensitive time-volume $\avgVTzero{}$ using a high \ac{FAR} (low
significance) threshold, the number of triggers with \acp{FAR} smaller
than one per century in subsequent experiments with sensitive
time-volume $\avgVT{}'$ will follow a Poisson distribution with mean
\begin{equation}
\label{eq:count-prime}
\Lambda' = \floud{} \Lambda \frac{\avgVT{}'}{\avgVTzero}.
\end{equation}
We plot the median value for $\Lambda'$ obtained in this way, as well
as the $90\%$ credible interval, as a function of surveyed time-volume
in the left panel of Figure~\ref{fig:lprime}.  There is,
unsurprisingly, a wide range of reasonable possibilities for the
number of highly significant events in future observations. The $90\%$
credible interval for the expected number of highly significant events
lies above one once $\avgVT{}'$ is approximately 1.5 times
$\avgVTzero$. As a point of reference, we show the expected value of
$\avgVT{}$ for the second and third planned observing runs, O2 and
O3. These volumes are calculated as in \citep{GW150914-ASTRO}, for an
equal-mass binary with non-spinning components and total mass
$60\,\Msun$, assuming an observation of 6 months for O2 and 9 months
for O3 with the same coincident duty cycle as during the first
\runtime{} of O1. We find estimates of
$\avgVT{}_{\mathrm{O2}}/\avgVTzero$ between \VTOtwolo{} and
\VTOtwohi{}, and $\avgVT{}_{\mathrm{O3}}/\avgVTzero$ between
\VTOthreelo{} and \VTOthreehi{}.  We also show $\avgVT{}$ for the
recently-completed O1 observing run, which surveyed approximately
three times the spacetime volume discussed here.  A paper describing
rate estimates using this methodology from the full O1 \ac{BBH} search
is in preparation.

Conditional on the count of loud events, $\Lambda'$, we can compute
the probability of having more than $n$ high-significance events in a
subsequent observation:
\begin{equation}
  \label{eq:pN-conditional}
  p\left( N > n | \Lambda' \right) = \exp\left[ - \Lambda' \right] \sum_{k = n+1}^\infty \frac{\Lambda'^k}{k!}.
\end{equation}
Applying Eq.\ \eqref{eq:count-prime}, and integrating over our
posterior on $\Lambda$ from the analysis in Section \ref{sec:counts},
we obtain the posterior probability of more than $n$ high-significance
events in a subsequent observation with sensitivity $\avgVT{}'$ given
our current observations:
\begin{multline}
  \label{eq:pN-posterior}
  p\left( N > n | \left\{ x_j \right\}, \avgVT{}' \right) = \\ \int
  \dd \Lambda_1 \, p\left( N > n | \Lambda'\left(\Lambda_1, \avgVT{}' \right)
  \right) p\left( \Lambda_1 | \left\{ x_j \right\}\right).
\end{multline}
The right panel of Figure \ref{fig:lprime} shows this probability for various values of
$n$ and $\avgVT{}'$.

The rates presented here are consistent with the theoretical
expectations detailed in \citet{Abadie2010}, but rule out the lowest
theoretically-allowed rates.  See \citet{GW150914-ASTRO} for a
detailed discussion of the implications of our rate estimates for
models of the binary \ac{BH} population.

\firstevent{} is unusually significant; only $\sim 8\%$ of the
astrophysical distribution of sources appearing in our search with a
threshold at \acp{FAR} of one per century will be more significant
than \firstevent{}.  However, it is not so significant as to call into
question the assumption used here that \ac{BBH} coalescences are
distributed uniformly in comoving volume and source time.  As we
accumulate more \ac{BBH} sources with ongoing Advanced LIGO observing
runs, we will eventually be able to test this assumption.  Similarly,
as we accumulate more sources and observation time, we will learn more
about the mass distribution of \ac{BBH} systems.  This is only the
beginning.

\acknowledgments

The authors gratefully acknowledge the support of the United States
National Science Foundation (NSF) for the construction and operation of the
LIGO Laboratory and Advanced LIGO as well as the Science and Technology Facilities Council (STFC) of the
United Kingdom, the Max-Planck-Society (MPS), and the State of
Niedersachsen/Germany for support of the construction of Advanced LIGO 
and construction and operation of the GEO600 detector. 
Additional support for Advanced LIGO was provided by the Australian Research Council.
The authors gratefully acknowledge the Italian Istituto Nazionale di Fisica Nucleare (INFN),  
the French Centre National de la Recherche Scientifique (CNRS) and
the Foundation for Fundamental Research on Matter supported by the Netherlands Organisation for Scientific Research, 
for the construction and operation of the Virgo detector
and the creation and support  of the EGO consortium. 
The authors also gratefully acknowledge research support from these agencies as well as by 
the Council of Scientific and Industrial Research of India, 
Department of Science and Technology, India,
Science \& Engineering Research Board (SERB), India,
Ministry of Human Resource Development, India,
the Spanish Ministerio de Econom\'ia y Competitividad,
the Conselleria d'Economia i Competitivitat and Conselleria d'Educaci\'o, Cultura i Universitats of the Govern de les Illes Balears,
the National Science Centre of Poland,
the European Commission,
the Royal Society, 
the Scottish Funding Council, 
the Scottish Universities Physics Alliance, 
the Hungarian Scientific Research Fund (OTKA),
the Lyon Institute of Origins (LIO),
the National Research Foundation of Korea,
Industry Canada and the Province of Ontario through the Ministry of Economic Development and Innovation, 
the Natural Science and Engineering Research Council Canada,
Canadian Institute for Advanced Research,
the Brazilian Ministry of Science, Technology, and Innovation,
Russian Foundation for Basic Research,
the Leverhulme Trust, 
the Research Corporation, 
Ministry of Science and Technology (MOST), Taiwan
and
the Kavli Foundation.
The authors gratefully acknowledge the support of the NSF, STFC, MPS, INFN, CNRS and the
State of Niedersachsen/Germany for provision of computational resources.
This article has been assigned the document number \href{https://dcc.ligo.org/LIGO-P1500217/public/main}{LIGO-P1500217}.

\bibliographystyle{aasjournal}
\bibliography{LIGO-P1500217_GW150914_Rates,../macros/GW150914_refs}

\begin{thebibliography}{}
\expandafter\ifx\csname natexlab\endcsname\relax\def\natexlab#1{#1}\fi

\bibitem[{{Aasi} {et~al.}(2013){Aasi}, {Abadie}, {Abbott}, {Abbott}, {Abbott},
  {Abernathy}, {Accadia}, {Acernese}, {Adams}, {Adams}, \& et~al.}]{Aasi2013}
{Aasi}, J., {Abadie}, J., {Abbott}, B.~P., {et~al.} 2013, \prd, 87, 022002

\bibitem[{{Abadie} {et~al.}(2010){Abadie}, {Abbott}, {Abbott}, {Abernathy},
  {Accadia}, {Acernese}, {Adams}, {Adhikari}, {Ajith}, {Allen}, \&
  et~al.}]{Abadie2010}
{Abadie}, J., {Abbott}, B.~P., {Abbott}, R., {et~al.} 2010, Classical and
  Quantum Gravity, 27, 173001

\bibitem[{Abbott {et~al.}(2016{\natexlab{a}})Abbott, Abbott, Abbott,
  {et~al.}}]{GW150914-ASTRO}
Abbott, B.~P., Abbott, R., Abbott, T.~D., {et~al.} 2016{\natexlab{a}},
  Astrophys.~J.~Lett., 818, L22

\bibitem[{Abbott {et~al.}(2016{\natexlab{b}})Abbott, Abbott, Abbott,
  {et~al.}}]{GW150914-CALIBRATION}
---. 2016{\natexlab{b}}, arXiv:1602.03845,
  \url{https://dcc.ligo.org/LIGO-P1500248/public/main}

\bibitem[{Abbott {et~al.}(2016{\natexlab{c}})Abbott, Abbott, Abbott,
  {et~al.}}]{GW150914-CBC}
---. 2016{\natexlab{c}}, Phys.~Rev.~D, 93, 122003

\bibitem[{Abbott {et~al.}(2016{\natexlab{d}})Abbott, Abbott, Abbott,
  {et~al.}}]{GW150914-DETECTION}
---. 2016{\natexlab{d}}, Phys.~Rev.~Lett., 116, 061102

\bibitem[{Abbott {et~al.}(2016{\natexlab{e}})Abbott, Abbott, Abbott,
  {et~al.}}]{GW150914-PARAMESTIM}
---. 2016{\natexlab{e}}, Phys.~Rev.~Lett., 116, 241102

\bibitem[{{Abbott} {et~al.}(2016){Abbott}, {Abbott}, {Abbott}, {Abernathy},
  {Acernese}, {Ackley}, {Adams}, {Adams}, {Addesso}, {Adhikari}, \&
  et~al.}]{Abbott2016GW151226}
{Abbott}, B.~P., {Abbott}, R., {Abbott}, T.~D., {et~al.} 2016, Physical Review
  Letters, 116, 241103

\bibitem[{Abbott {et~al.}(2016)}]{RatesSupplement}
Abbott, B.~P., {et~al.} 2016, \apjs, TBD, TBD

\bibitem[{Allen {et~al.}(2012)Allen, Anderson, Brady, Brown, \&
  Creighton}]{Allen:2005fk}
Allen, B., Anderson, W.~G., Brady, P.~R., Brown, D.~A., \& Creighton, J. D.~E.
  2012, Phys. Rev., D85, 122006

\bibitem[{{Chen} \& {Holz}(2014)}]{Chen2014}
{Chen}, H.-Y., \& {Holz}, D.~E. 2014, ArXiv e-prints, arXiv:1409.0522

\bibitem[{{Dominik} {et~al.}(2012){Dominik}, {Belczynski}, {Fryer}, {Holz},
  {Berti}, {Bulik}, {Mandel}, \& {O'Shaughnessy}}]{Dominik2012}
{Dominik}, M., {Belczynski}, K., {Fryer}, C., {et~al.} 2012, \apj, 759, 52

\bibitem[{{Farr} {et~al.}(2015){Farr}, {Gair}, {Mandel}, \&
  {Cutler}}]{Farr2015}
{Farr}, W.~M., {Gair}, J.~R., {Mandel}, I., \& {Cutler}, C. 2015, \prd, 91,
  023005

\bibitem[{{Farr} {et~al.}(2011){Farr}, {Sravan}, {Cantrell}, {Kreidberg},
  {Bailyn}, {Mandel}, \& {Kalogera}}]{Farr2011}
{Farr}, W.~M., {Sravan}, N., {Cantrell}, A., {et~al.} 2011, \apj, 741, 103

\bibitem[{{Fryer} {et~al.}(2012){Fryer}, {Belczynski}, {Wiktorowicz},
  {Dominik}, {Kalogera}, \& {Holz}}]{Fryer2012}
{Fryer}, C.~L., {Belczynski}, K., {Wiktorowicz}, G., {et~al.} 2012, \apj, 749,
  91

\bibitem[{{Fryer} \& {Kalogera}(2001)}]{Fryer2001}
{Fryer}, C.~L., \& {Kalogera}, V. 2001, 554, 548

\bibitem[{{Gladman} {et~al.}(1998){Gladman}, {Kavelaars}, {Nicholson},
  {Loredo}, \& {Burns}}]{Gladman1998}
{Gladman}, B., {Kavelaars}, J.~J., {Nicholson}, P.~D., {Loredo}, T.~J., \&
  {Burns}, J.~A. 1998, \aj, 116, 2042

\bibitem[{{Guglielmetti} {et~al.}(2009){Guglielmetti}, {Fischer}, \&
  {Dose}}]{Guglielmetti2009}
{Guglielmetti}, F., {Fischer}, R., \& {Dose}, V. 2009, \mnras, 396, 165

\bibitem[{{Hogg}(1999)}]{Hogg1999}
{Hogg}, D.~W. 1999, ArXiv Astrophysics e-prints, arXiv:astro-ph/9905116

\bibitem[{{Kelly} {et~al.}(2009){Kelly}, {Vestergaard}, \& {Fan}}]{Kelly2009}
{Kelly}, B.~C., {Vestergaard}, M., \& {Fan}, X. 2009, \apj, 692, 1388

\bibitem[{{Kim} {et~al.}(2003){Kim}, {Kalogera}, \& {Lorimer}}]{Kim2003}
{Kim}, C., {Kalogera}, V., \& {Lorimer}, D.~R. 2003, \apj, 584, 985

\bibitem[{{Kreidberg} {et~al.}(2012){Kreidberg}, {Bailyn}, {Farr}, \&
  {Kalogera}}]{Kreidberg2012}
{Kreidberg}, L., {Bailyn}, C.~D., {Farr}, W.~M., \& {Kalogera}, V. 2012, \apj,
  757, 36

\bibitem[{{Littenberg} {et~al.}(2013){Littenberg}, {Baker}, {Buonanno}, \&
  {Kelly}}]{Littenberg2013}
{Littenberg}, T.~B., {Baker}, J.~G., {Buonanno}, A., \& {Kelly}, B.~J. 2013,
  \prd, 87, 104003

\bibitem[{{Loredo} \& {Wasserman}(1995)}]{Loredo1995}
{Loredo}, T.~J., \& {Wasserman}, I.~M. 1995, \apjs, 96, 261

\bibitem[{{Loredo} \& {Wasserman}(1998{\natexlab{a}})}]{Loredo1998Iso}
---. 1998{\natexlab{a}}, \apj, 502, 75

\bibitem[{{Loredo} \& {Wasserman}(1998{\natexlab{b}})}]{Loredo1998Aniso}
---. 1998{\natexlab{b}}, \apj, 502, 108

\bibitem[{{Messick} {et~al.}(2016){Messick}, {Blackburn}, {Brady}, {Brockill},
  {Cannon}, {Caudill}, {Chamberlin}, {Creighton}, {Everett}, {Hanna}, {Lang},
  {Li}, {Meacher}, {Pankow}, {Privitera}, {Qi}, {Sachdev}, {Sadeghian},
  {Sathyaprakash}, {Singer}, {Thomas}, {Wade}, {Wade}, \&
  {Weinstein}}]{gstlalmethods:2015}
{Messick}, C., {Blackburn}, K., {Brady}, P., {et~al.} 2016, ArXiv e-prints,
  arXiv:1604.04324

\bibitem[{{{\"O}zel} {et~al.}(2010){{\"O}zel}, {Psaltis}, {Narayan}, \&
  {McClintock}}]{Ozel2010}
{{\"O}zel}, F., {Psaltis}, D., {Narayan}, R., \& {McClintock}, J.~E. 2010,
  \apj, 725, 1918

\bibitem[{{Petit} {et~al.}(2008){Petit}, {Kavelaars}, {Gladman}, \&
  {Loredo}}]{Petit2008}
{Petit}, J.-M., {Kavelaars}, J.~J., {Gladman}, B., \& {Loredo}, T. 2008, {Size
  Distribution of Multikilometer Transneptunian Objects}, ed. M.~A. {Barucci},
  H.~{Boehnhardt}, D.~P. {Cruikshank}, A.~{Morbidelli}, \& R.~{Dotson}, 71--87

\bibitem[{{Planck Collaboration} {et~al.}(2015){Planck Collaboration}, {Ade},
  {Aghanim}, {Arnaud}, {Ashdown}, {Aumont}, {Baccigalupi}, {Banday},
  {Barreiro}, {Bartlett}, \& et~al.}]{Planck2015}
{Planck Collaboration}, {Ade}, P.~A.~R., {Aghanim}, N., {et~al.} 2015, ArXiv
  e-prints, arXiv:1502.01589

\bibitem[{{P{\"u}rrer}(2015)}]{Purrer2015}
{P{\"u}rrer}, M. 2015, ArXiv e-prints, arXiv:1512.02248

\bibitem[{{Salpeter}(1955)}]{Salpeter1955}
{Salpeter}, E.~E. 1955, \apj, 121, 161

\bibitem[{{Schutz}(2011)}]{Schutz2011}
{Schutz}, B.~F. 2011, Classical and Quantum Gravity, 28, 125023

\bibitem[{{Spera} {et~al.}(2015){Spera}, {Mapelli}, \& {Bressan}}]{Spera2015}
{Spera}, M., {Mapelli}, M., \& {Bressan}, A. 2015, \mnras, 451, 4086

\bibitem[{Taracchini {et~al.}(2014)}]{Taracchini:2013rva}
Taracchini, A., {et~al.} 2014, Phys. Rev., D89, 061502

\bibitem[{{The LIGO Scientific Collaboration} {et~al.}(2016){The LIGO
  Scientific Collaboration}, {the Virgo Collaboration}, {Abbott}, {Abbott},
  {Abbott}, {Abernathy}, {Acernese}, {Ackley}, {Adams}, {Adams}, \&
  et~al.}]{Abbott2016O1BBH}
{The LIGO Scientific Collaboration}, {the Virgo Collaboration}, {Abbott},
  B.~P., {et~al.} 2016, ArXiv e-prints, arXiv:1606.04856

\bibitem[{{Usman} {et~al.}(2015){Usman}, {Kehl}, {Nitz}, {Harry}, {Brown},
  {Capano}, {Dent}, {Fairhurst}, {Pfeiffer}, {Biwer}, {Dal Canton}, {Keppel},
  {Saulson}, {West}, \& {Willis}}]{Usman2015}
{Usman}, S.~A., {Kehl}, M.~S., {Nitz}, A.~H., {et~al.} 2015, ArXiv e-prints,
  arXiv:1508.02357

\bibitem[{{Veitch} {et~al.}(2015){Veitch}, {Raymond}, {Farr}, {Farr}, {Graff},
  {Vitale}, {Aylott}, {Blackburn}, {Christensen}, {Coughlin}, {Del Pozzo},
  {Feroz}, {Gair}, {Haster}, {Kalogera}, {Littenberg}, {Mandel},
  {O'Shaughnessy}, {Pitkin}, {Rodriguez}, {R{\"o}ver}, {Sidery}, {Smith}, {Van
  Der Sluys}, {Vecchio}, {Vousden}, \& {Wade}}]{Veitch2015}
{Veitch}, J., {Raymond}, V., {Farr}, B., {et~al.} 2015, \prd, 91, 042003

\end{thebibliography}

\allauthors

\end{document}